# The Image of the City Out of the Underlying Scaling of City Artifacts or Locations

Bin Jiang
Department of Technology and Built Environment, Division of Geomatics
University of Gävle, SE-801 76 Gävle, Sweden
Email: bin.jiang@hig.se

**Abstract**
Two fundamental issues surrounding research on the image of the city respectively focus on the city's external and internal representations. The external representation in the context of this paper refers to the city itself, external to human minds, while the internal representation concerns how the city is represented in human minds internally. This paper deals with the first issue, i.e., what trait the city has that make it imageable? We develop an argument that the image of the city arises from the underlying scaling of city artifacts or locations. This scaling refers to the fact that, in an imageable city (a city that can easily be imaged in human minds), small city artifacts are far more common than large ones; or alternatively low dense locations are far more common than high dense locations. The sizes of city artifacts in a rank-size plot exhibit a heavy tailed distribution consisting of the head, which is composed of a minority of unique artifacts (vital and very important), and the tail, which is composed of redundant other artifacts (trivial and less important). Eventually, those extremely unique and vital artifacts in the top head or those largest so to speak, i.e., what Lynch called city elements, make up the image of the city. We argue that the ever-increasing amount of geographic information on cities, in particular obtained from social media such as Flickr and Twitter, can turn research on the image of the city, or cognitive mapping in general, into a quantitative manner. The scaling property might be formulated as a law of geography.

**1. Introduction**
There are two fundamental issues surrounding research on cognitive mapping or the image of the city in particular. The first refers to the city's external representation, the city itself external to human minds, while the second concerns how the city is internally represented in human minds. Cognitive mapping is the mental process of producing mental or cognitive maps or images in human minds about our physical environments or cities. In this context, a cognitive map is a mental image or internal representation of the real world (Downs and Stea 1973). Relying on human subjects as experimental instruments such as questionnaire surveys, interviews and sketch maps, Lynch (1960) developed two concepts to formulate the theory on the image of the city. The first concept is legibility. A legible city can give an observer a clear and distinct layout, thus easily shaping the image of the city in human minds. By contrast, an illegible city can hardly be processed to form a mental map, and the city's layout is little clear to an observer. The second related concept is imageability. It is a quality of cities or city artifacts that gives people a strong and vivid image. Cities with imageable artifacts tend to be imageable or legible, and legible cities contain vivid imageable artifacts or elements that make up the image of the city.

Lynch attempted to characterize cognitive mapping from the perspective of cities and city artifacts, the external representation, rather than that of human brains or minds, the internal representation. Following this work, fruitful research (e.g., Kitchin and Freundschuh 2000, Portugali 1996) has been performed in the domain of cognitive mapping, with a particular focus on the nature of internal representation and how the internal representation may impact the behavior of individuals (Gärling and Golledge 1993, Golledge and Stimson 1997). The increasing availability of geographic information has made cognitive mapping or spatial cognition an interesting research topic in the emerging field of geographic information science (Montello 2002, Omer et al. 2005, Montello 2009). However, little research over the past decades has been done on the external representation by addressing questions such as these: Why can the image of the city be formed? What cities are legible? What city artifacts are imageable? Can we compare two cities in terms of legibility and imageability? Ultimately, why can the image of the city be formed from the perspective of external representation?

In seeking an answer to why the image of the city can be formed, Haken and Portugali (2003) developed an information theory perspective that focuses on the external face of the city. They argued and demonstrated that legible cities contain more information than illegible ones, and imageable city artifacts contain more information than unimageable artifacts. They drew related evidence in the fields of information theory and synergetics (Shannon and Weaver 1949, Haken 1983) to support this argument. In particular, they placed a considerable emphasis on semantic information based on Haken's synergetics to differ from Shannonian objective information, which lacks any attached semantic meaning. Similar to Lynch's original study, this study concentrates on the nature and role of external representation, but leaves out the internal representation. This is an important development toward addressing the questions that were raised above.

In this paper, we develop an alternative yet more intuitive view that the face or the image of the city is attributed to the underlying scaling of city artifacts or locations. Scaling is a well-studied phenomenon in a variety of disciplines, including for instance physics, mathematics, economics, linguistics, biology and computer sciences (e.g., Barenblatt 2003, Clauset et al. 2009, and references therein). The scaling, in the context of geography, implies that there are more small things than large things, or far more low dense locations than high dense locations, within a geographic space. For example, there are far more small cities than large ones, which is a phenomenon characterized by Zipf's law (1949); far more small city artifacts than large city artifacts in a city (e.g., Salingaros 2005, Carvalho and Penn 2004, Jiang 2007, Jiang 2009); far large countryside than cities in terms of land occupied (Jiang and Liu 2012). We in this paper develop a quantitative approach to producing cognitive maps based on massive amounts of geographic information and by ranking importance or relevance of individual city artifacts. All these city artifacts are placed into two categories, i.e., unique and redundant artifacts, which correlate with the head and the tail, respectively, of a heavy-tailed distribution that is skewed far to the right in a rank-size plot. Eventually, the unique artifacts in the top head, or those largest, make up the image of the city.

The contribution of this paper is three-fold: (1) we uncover a fundamental mechanism of why the image of the city can be formed from the perspective of cities or city artifacts or that of the external representation; (2) we turn research on cognitive mapping into a quantitative manner, based on massive geographic information, as an alternative to existing methods using human beings as experimental instruments; and (3) we establish a link between cognitive mapping and cartographic mapping, and unify the two types of mapping under the same framework of scaling. Now that scaling is an essential property of geographic space, both cognitive maps and cartographic maps must capture the essence of the representation of geographic space. This is a central point we want to argue and advocate in this paper.

The remainder of this paper is structured as follows. Section 2 briefly reviews the theory of the image of the city, in particular the city elements, and the difference between individual and collective mental maps. Section 3 introduces the mathematics of scaling, which is characterized by various nonlinear relationships between a variable $x$ and its probability $y$. In Section 4, we illustrate that the image of the city arises from the underlying scaling by demonstrating that (1) a city with the underlying scaling property is alive rather than dead, (2) scaling is a form of deep order rather than surface order from both geometric and topological perspectives, and (3) both Manhattan and Paris with seemingly different morphologies have the same deep order from the topological perspective. In section 5, we further discuss the scaling view and its relationship to the information view (Haken and Portugali 2003), and place both cognitive mapping and cartographic mapping under the framework of scaling. Finally, we draw conclusions and outline future work.

**2. The image of the city**
A legible city usually contains five types of imageable elements: paths, edges, districts, nodes and landmarks (Lynch 1960). All of these city elements, despite having different geometric shapes or visual qualities, have one thing in common: they are all distinguished from redundant others and give a vivid image to an observer. These imageable city elements make a city legible and subsequently make up the image of the city. Because of their distinctiveness, the city elements are often referred to as landmarks in a broader sense (Haken and Portugali 2003). In general terms, the city elements *afford* their ability to be perceived, remembered and imagined. This is where the concepts of legibility and imageability are closely related to Gibson's (1979) affordances (Haken and Portugali 2003). An affordance is the quality of any artifact or environment that allows an individual to perform an action (e.g., books for reading, buttons for pushing, and streets for walking or driving). Note that in this paper, we deliberately take "city elements" or "landmarks" to be words reserved to refer to those city artifacts that make up the image of the city.

City artifacts that are qualified to be city elements or landmarks are not constrained to their distinguished visual or geometric qualities. In fact, any city artifacts that bear special meanings and are distinguished from redundant other artifacts are qualified to be landmarks. The special meanings can be cultural, historical, or personal. For example, a place that has no distinguished geometric shape or visual quality, but in which someone had a terrible traffic accident, tends to be a landmark for this particular person. If the place was associated with a traffic accident that killed a very important person (e.g., a queen of the country), it would be a landmark for all the city residents. A small house, in terms of its geometric or visual quality, is not qualified as a city element. However, it could be qualified if the house has a long history or was owned by a world-renowned person. Landmarks vary from one person to another, but some landmarks constitute the mental map for nearly all people, e.g., London Bridge and Eiffel Tower.

In a seminal work on eye movement and vision, Yarbus (1967) showed examples in which some parts of a picture attract far more attention than other parts. Figure 1 illustrates one such example for a human face. To the left in the figure is a photographed picture, and to the right are the tracked eye movement trajectories. The intensity of the fixation points shows the degree of attention by an observer. The recorded eye movements or the points of intensity reflect a mental image of the picture of the face. We can see that both eyes and the mouth receive much more attention than the other parts of the facial 'landscape' or other objects (such as the hair and the scarf) in the picture. Thus, both the eyes and the mouth constitute the landmarks of the face because they are distinguished from the redundant others (other parts or objects).

Note that we have two perspectives for the reading or perception of this face. The first is the spatial perspective, i.e., only a very small part of the face landscape or space is occupied by the eyes and mouth, whereas most of the other parts of the face remain unoccupied. The second is the perspective of recognizable objects. There are many objects, notably the eyes, nose, mouth, hair, and scarf, but only the eyes and mouth are highly perceived, whereas other objects are mostly neglected. This result appears to suggest that the large and complex objects usually capture more attention during the perception process. For example, largest cities receive far more attention than small ones, and cities receive more attention than the countryside because of the complexity of the cities.

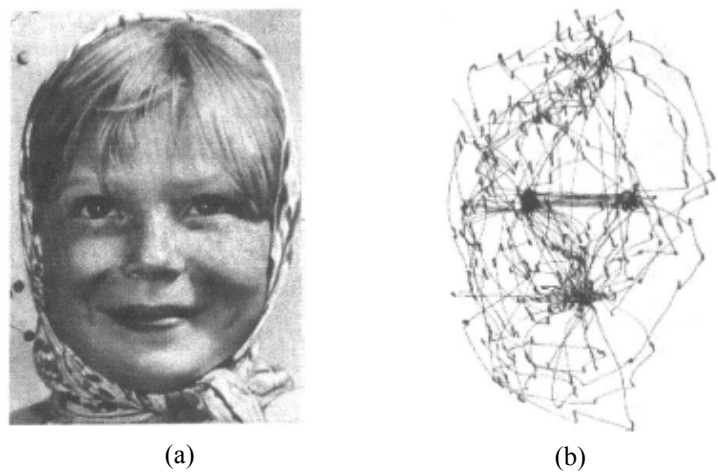

(a)　　　　　　　　　　　　　　(b)

Figure 1: Illustration of (a) a photographed human face, and (b) its eye movement trajectories (Yarbus 1967)

Mental map or the image of the city is a selective representation of individuals' own city in terms of how familiar the city artifacts or locations are and also in terms of the individuals' cultural and historical backgrounds. A mental map is closely related to one's daily life and business. For example, a taxi driver would have an entire image of the city street network, whereas a person who stays in one neighborhood most of the time is likely to have a clear image of the immediate surroundings but a limited idea of the entire city. A person who has lived in a city for a lifetime can have a detailed image of the city, whereas a tourist who just visits the city for a few days only has a small impression of limited locations. Therefore, mental maps are mostly personal and unique, and they involve a series of cognitive process, such as selection, simplification, categorization, deletion, distortion and generalization (Downs and Stea 1973). In addition to the direct experience, our mental maps are also significantly shaped by the indirect experience obtained through TV, newspapers, films and other media. However, this paper concentrates on a collective mental map, which is composed by landmarks or city elements nearly for all city residents. In this regard, the collective mental map can be considered to a wisdom created by crowds (Surowiecki 2005). We developed an argument that the collective mental map or the image of the city arises from the scaling of city artifacts or locations. Thus the image of the city is computable based on massive geographic information (Jiang 2012b).

# 3. Mathematical models for characterizing scaling

In this section, we introduce some mathematical models that are used to characterize the scaling property. The mathematical models are also called heavy-tailed distributions, which are skewed to the right. The skewed distributions represent a sharp contrast to a normal distribution, in which all values of a variable center around an average value or mean. By contrast, the heavy-tailed distributions have no meaningful average value that characterizes the individual values. To illustrate the difference between a heavy-tailed distribution and a normal distribution, we generated two sets of random numbers. The first set follows a normal distribution, whereas the second exhibits a heavy-tailed distribution. We classified the two datasets, and the resulting classes are shown in Table 1. Note that the classification for the normal distribution dataset is based on Jenks' (1967) natural break, while the classification for the heavy-tailed distribution is based on head/tail breaks (Jiang 2012a). This is because the head/tail breaks classification captures the underlying scaling or hierarchy of a heavy-tailed distribution.

Table 1: Classes or hierarchical levels of the two generated datasets
(Note: The left half shows the normal distribution, and the right half shows the heavy-tailed distribution)

| Class | Size | Freq | % | Class | Size | Freq | % |
|---|---|---|---|---|---|---|---|
| 1 | 44 | 69 | 6.90% | 1 | 56 | 823 | 82.30% |
| 2 | 50 | 191 | 19.10% | 2 | 216 | 135 | 13.50% |
| 3 | 56 | 278 | 27.80% | 3 | 568 | 28 | 2.80% |
| 4 | 61 | 245 | 24.50% | 4 | 1049 | 9 | 0.90% |
| 5 | 67 | 167 | 16.70% | 5 | 1659 | 4 | 0.40% |
| 6 | 81 | 50 | 5.00% | 6 | 2597 | 1 | 0.10% |

If we plot histograms of the datasets, we clearly see the difference between the two distributions (Figure 2). The normal distribution has a bell-shaped form, whereas the heavy-tailed distribution is skewed to the right. With the normal distribution, the sizes of small and large classes are almost equal because the symmetric bell shape. With the heavy-tailed distribution, the size of the small class is 82.30%, whereas that of the large classes is just $1-82.3\% = 17.7\%$. Thus, there is an imbalance between the sizes of the small and large classes. This is a distinguishing feature for any heavy-tailed distribution. Both plots shown in Figure 2 are histograms. To introduce the heavy-tailed distributions that are consistent with head/tail breaks (Jiang 2012a), we must introduce the rank-size distribution.

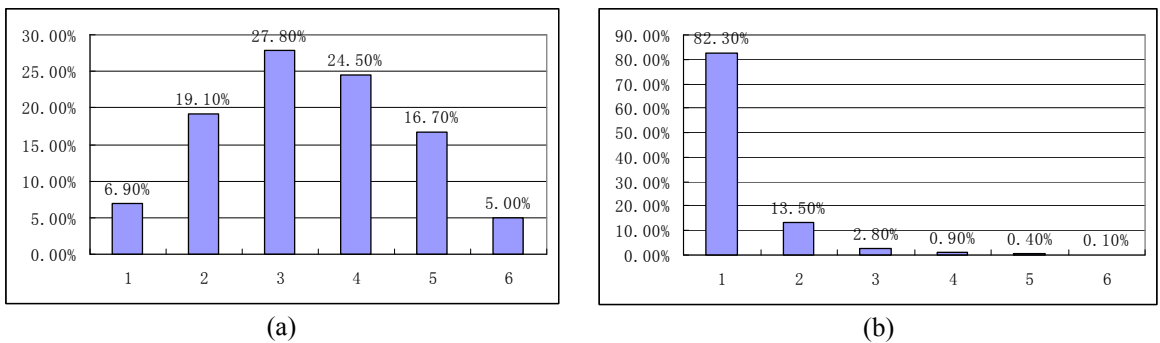

Figure 2: Histograms of (a) the normal distribution and (b) the heavy-tailed distribution
(Note: The x-axis represents the classes, and the y-axis represents the corresponding frequency in percentages)

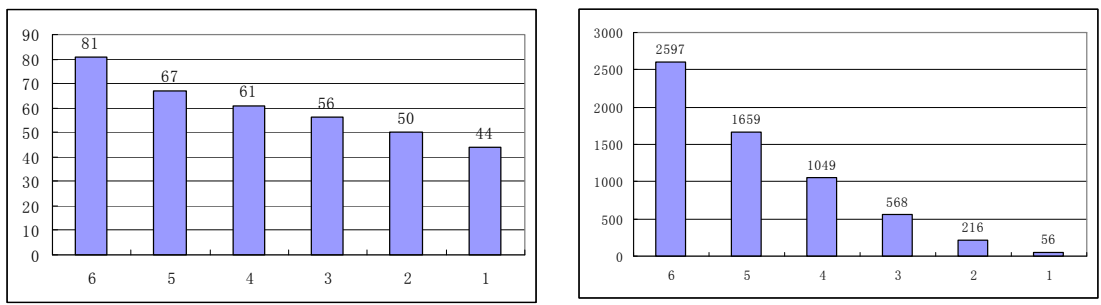

(a) (b)

Figure 3: Rank-size plots of (a) the normal distribution and (b) the heavy-tailed distribution
(Note: The x-axis represents the class number ranked from the largest class to the smallest class. In other words, class 7 is ranked number 1, followed by class 6 as number 2, and so on.)

A rank-size distribution ranks the classes or values of a variable in a decreasing order from the left to the right, whereas the y-axis shows the size or frequency. This rank-size distribution is also called Zipf's law (Zipf 1949) for the analysis of city size distribution and word frequency distribution. For the dataset that exhibits the heavy-tailed distribution (i.e., the right half of Table 1), class 6 is ranked as number 1, followed by class 5, class 4, and so on (Figure 3b). This rank-size distribution, unlike the normal distribution (Figure 3a), still looks heavy-tailed and shows large classes in the head and small classes in the tail. Importantly, the large classes in the head occupy a small percentage, whereas the small classes in the tail occupy a large percentage (82.30%). For the sake of simplicity and for a better understanding, we plot the heavy-tailed distribution in a discrete mode and using a limited number of classes (6). For the sake of good mathematics, we will briefly introduce three heavy-tailed distributions in the form of continuous functions.

In the literature, especially in the literature of statistical physics, scaling refers to phenomena with a power law distribution. However, we in this paper relax this notion of scaling to include three heavy-tailed distributions: power law, lognormal and exponential. Strictly speaking, heavy-tailed distributions are defined as distributions whose tails are heavier or longer than the exponential distribution, thereby excluding exponential distributions from being heavy-tailed. Given $x$ the values of a variable (e.g., street lengths or city sizes) and $y$ the probability of $x$, there are three possible relationships between $x$ and $y$, which are heavy tailed.

The first is the power law distribution, which is expressed by

$$y = x^{-\alpha}, \qquad [1]$$

where α is called the power law exponent.

If we take the logarithm for equation [1], we obtain

$$\ln y = -\alpha \ln x. \qquad [2]$$

It is clear that the distribution line is straight in a double logarithm plot. The distribution line is often used to detect the power law through the double logarithm plot. However, this method has been criticized for creating errors or bias when drawing a conclusion on power law distributions. Instead, some methods based on maximum likelihood have been suggested (Clauset et al. 2009) to detect a power law distribution.

The second formula is the lognormal function, which has the following format:

$$y = \frac{1}{\sqrt{2\pi\sigma^2}\,x} \exp(\frac{-(\ln(x)-\mu)^2}{2\sigma^2}). \qquad [3]$$

The third formula is the exponential function, which is expressed by

$$y = e^x. \qquad [4]$$

If we take the logarithm for equation [4], we will obtain

$$\ln y = x. \qquad [5]$$

Equation [5] indicates that x and the logarithm of y have a linear relationship. Note that the constant $e$ can be any other constant value (e.g., $y = C^x$), where the constant C can be set to $\frac{1}{3}$ for example.

The above three distributions with an appropriate parameter setting can be very similar, with a long tail or heavy tail. We have tried to keep the mathematics at the simplest level. The above formulae are therefore presented in a simple format by ignoring some constants or parameters. Interested readers may refer to the literature (e.g., Clauset et al. 2009 and references therein) for more detailed descriptions and elaborations. Apart from the three formulae, their degenerate versions such as the power law with an exponential cutoff and stretched exponential distribution are often observed with real-world data.

**4. Case studies**
The tool used for the following case studies is the least number of the longest lines, or the axial lines that cut across the free space of an urban environment, and close all loops, initially developed in space syntax (Hillier and Hanson 1984, Hillier 1996). Note that the axial lines are conventionally drawn manually by following some rigid rule, i.e., starting with the first longest line, followed by the second longest, the third longest and so on until the free space is fully covered by the intersecting lines. However, the axial lines or the axial maps used in the case studies are automatically generated using software Axwoman (http://fromto.hig.se/~bjg/Axwoman/). These axial lines are not city artifacts per se, but they bear the same scaling property as city artifacts. This justifies their use in the case studies. More importantly, the axial lines capture the underlying morphological structure of cities. Therefore, the case studies are based on two basic assumptions: (1) all lines are treated as all city artifacts of individual cities; no other city artifacts are left out on the maps or in the cities, and (2) no semantic information is available, which means that only geometric and topological information is available for generating the mental maps. The reason why we add these two assumptions is that we want to show for an illustration purpose how mental maps can be generated from the city artifacts (or the axial lines), and governed by the underlying scaling of the city artifacts or locations.

**4.1 Dead city versus living city**
For the sake of simplicity, we adopt two small urban environments (bear with me for calling them cities, since they are too small to be cities), one artificial and the other real, to illustrate the concepts of dead and living cities (Figure 4). The artificial city has four rectangle street blocks and an outmost rectangular border, whereas the real city (namely Gassin taken from Hillier and Hanson 1984) has a much more convoluted shape of blocks and an outmost border. We call the artificial city a dead city because it lacks of change in its form. The 7 axial lines have almost the same length and range from 1448 to 1882 units. The ratio of the maximum to the minimum is 1882/1448=1.3, which is a very small variance.

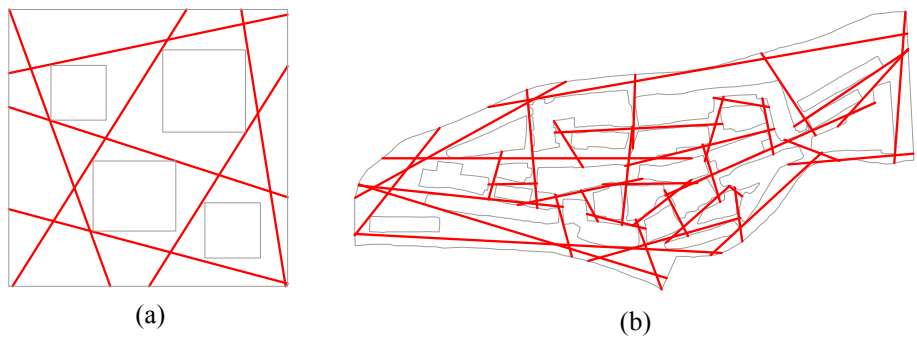

Figure 4: (Color online) Illustration of (1) a dead city, and (b) a living city

On the other hand, the real city is characterized by many changes in terms of the shape of the street blocks and the free space between the blocks. The free space is covered by the 39 axial lines, with all lengths ranging from 76 to 1896 units. The ratio of the maximum to the minimum is 1896/76=24.9, which is very big in particular compared to the artificial city. As noted above, the smallest set of the axial lines represents how the free space is perceived by individuals as fragmented, yet intersected, linear spaces. In fact, the lengths of the streets for the dead and living city follow respectively a normal distribution and a heavy-tailed distribution. The real city has many geometric changes, whereas the artificial city lacks of changes. The same observation can be made from a topological perspective, i.e., the degree of connectivity of individual lines. The connectivity for the artificial city is 3 or 4, while the connectivity for the real city varies from 2 to 39, so a wide range.

The concepts of dead and living cities are closely related to that of living structure (Alexander 2004). We believe that the scaling property underlies the living structure or living city. In this regard, Salingaros (2005) formulated three laws of architecture that concern orders on the small scale, the large scale and intermediate scales. All of the architectural constituents or objects at the three identifiable scales form a hierarchical structure as a whole. Although initially developed for architecture, the three laws appear to be valid for urban environments or cities as well.

## 4.2 Surface order versus deep order

This case study use some typical street patterns in the literature (Jacobs 1995) and examine how the free space between street blocks might emerge and be perceived in human minds. A spectrum of color is adopted for visualization of the city structure, with red and blue representing the largest and smallest spaces respectively; the other colors represent the spaces in between (Figure 5). The classification of spaces was performed according to head/tail breaks (Jiang 2012a), because it is able to capture the underlying scaling property of axial lines. Note that the coloring indicates relative values within a city, and it cannot be used for cross comparisons between different cities. For example, the longest streets in Copenhagen are not the same length as the longest streets in London. Relying on the coloring, we examine how the images of the street patterns may be shaped in human minds.

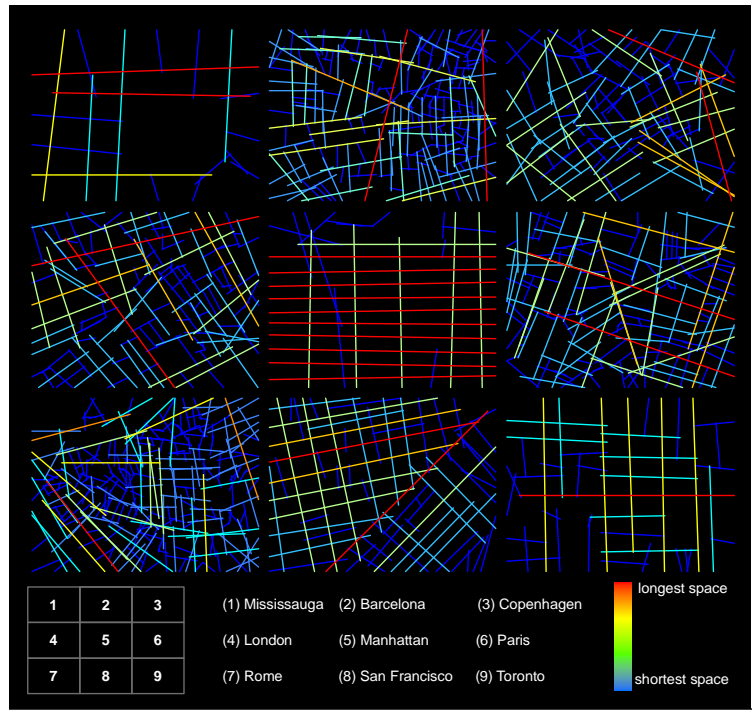

Figure 5: (Color online) Geometry of the typical street patterns

North American cities are distinct from European cities, with the former being regular and planned and the latter being irregular and self-evolving. North American cities demonstrate a grid-like form, and most of the streets are oriented north-south or east-west. The main streets, such as those of San Francisco, are parallel to each other and exhibit a grid-like form. If we further examine the streets, we find that there are very few scales (or sizes) involved. The most extreme case is Manhattan. Almost all streets in Manhattan are very long and the same length, and there are few short streets. More critically, Manhattan lacks of a full range of intermediate scales (or sizes). In other words, the linkage between scales is missing (Salingaros 2005). In this regard, the other three North American cities are not as simple as Manhattan in terms of overall structure.

By contrast, the geometry of the European cities looks random or arbitrary but shows a living structure with a full range of scales of length and orientation. In general, we can place the five European cities into two groups: the first group (Barcelona, London and Paris) and the second group (Copenhagen and Rome). The reason for this grouping is that the latter cities, unlike the former, lack of the longest streets that cross the entire region. In other words, both Copenhagen and Rome miss the largest scale that crosses the entire area. Note that the above remarks about living or dead structures are derived from the geometric aspect of the street length only and for the sampled areas only. In reality, peoples' reactions about dead or living cities would include other factors, such as trees, shops, and the geometries of facades.

Having illustrated the differences between the forms of North American and European cities in their forms, we now define the concepts of surface order and deep order. By surface order, it means rectilinear geometry, such as rectangular shape and west-east or north-south orientation (as shown in most North American cities). Most European cities show a type of deep order that has a full range of lengths and orientations. The surface order is more intuitive and more perceivable than the deep order. Most people tend to confuse the deep order with disorder or regard it as being arbitrary. Deep order is not disorder; rather, it is a sort of hidden order (Holland 1995, Hanson 1989). The hidden order can be characterized as a needle-in-a-haystack type of order. A needle in a haystack is used to describe a situation in which something is difficult to be distinguished or to be found. However, a needle in a haystack in a digital environment is easy to find, given that semantic information is attached to the needle and the hay. The imbalance between a tiny needle and a vast majority of hay characterizes the difference between the needle and the hay.

**4.3 From geometry to topology**
Accumulating evidence suggests that human perception of the city or physical environments is generally topological rather than geometric (e.g., Penn 2003). This is evident in the space syntax research showing that topological measures can be a good indicator for human traffic flow (Hillier 1996). In what follows, we adopt street connectivity to replace street length. Although street length tends to correlate with street connectivity, geometric length has never been a predictor for traffic flow. A partial reason for this is that geometric lengths are not a major factor in spatial cognition. A well connected street tends to capture more attention than a long street. This can be easily understood by the fact that a well connected street is in a city, and a long street that is not a well connected street is usually in the countryside. Figure 6 illustrates the topological patterns based on the degree of street connectivity. We can see that the underlying patterns are indeed different from what was illustrated by street length, as shown in Figure 5. Overall, there are fewer most-connected streets (red ones in Figure 6) than longest streets (red lines in Figure 5). Red lines or the most-connected ones tend to form part of the image of the city. The red lines are unique, whereas the others are redundant. In other words, the red lines indicate the highest degree of legibility or imageability, whereas the other lines constitute the background or context.

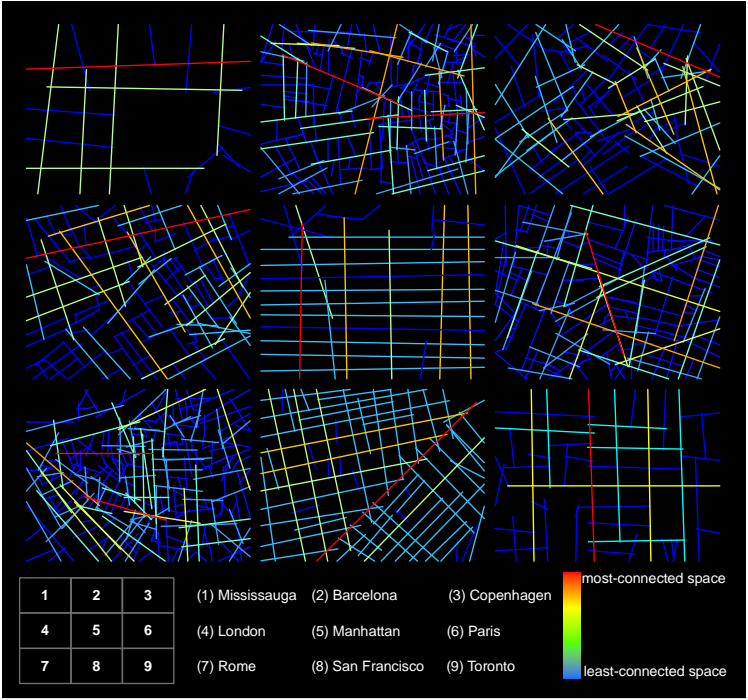

Figure 6: (Color online) Topology of the typical street patterns

It is important to note that the above discussions on various cities have some biases because they are not sufficiently large and are just parts of the large cities used for an illustration purpose. Next, we will illustrate that two seemingly different morphologies, i.e., Manhattan and Paris, in fact have the same deep order. Both cities are sufficiently large and complete for a further examination from the perspective of topology, or in terms of street connectivity. Manhattan has 1840 streets, whereas Paris has 6846. There are no isolated streets or lines involved, so the lowest connectivity is 1 for both cities. The degree of the most-connected streets varies from one city to another: 186 for Manhattan and 77 for Paris. These differences do not cover the deep order that is to be further examined and illustrated.

We found that Manhattan and Paris have respectively 7 and 9 hierarchical levels or classes. The result is shown in Table 2, in which classes, sizes, frequency and percentages are shown for the two cities. The classes are equivalent to the ranking of streets in a decreasing order from the most-connected to the least-connected. The only difference between the classes and the rank is the range or intervals: the classes range from 1-7, whereas the rank ranges from 1-1840 for Manhattan (because there are 1840 streets). Table 2 indicates that there are far more less-connected streets than well-connected ones, which is further illustrated in Figure 7, where there are far more blue lines than red ones. Importantly, there are several intermediate scales from the least connected to the most connected. What is most interesting and surprising to us is that both cities have almost identical distributions of streets in the classes, even though they have seemingly different morphologies, and have different hierarchal levels or classes.

Table 2: Classes or hierarchical levels of the streets of Manhattan (the left half) and Paris (the right half)

| Class | Size | Freq | % | Class | Size | Freq | % |
|---|---|---|---|---|---|---|---|
| 1 | 187 | 2 | 0.11% | 1 | 77 | 1 | 0.01% |
| 2 | 144 | 3 | 0.16% | 2 | 63 | 2 | 0.03% |
| 3 | 122 | 6 | 0.33% | 3 | 54 | 6 | 0.09% |
| 4 | 80 | 15 | 0.82% | 4 | 44 | 18 | 0.26% |
| 5 | 33 | 78 | 4.24% | 5 | 34 | 43 | 0.63% |
| 6 | 16 | 428 | 23.26% | 6 | 26 | 153 | 2.23% |
| 7 | 7 | 1308 | 71.09% | 7 | 17 | 483 | 7.06% |
| NOTE: Manhattan has 7 classes, | | | | 8 | 10 | 1092 | 15.95% |
| while Paris has 9 classes. | | | | 9 | 6 | 5048 | 73.74% |

The minority of the well-connected streets or the most-connected ones constitute the mental maps. There are only two streets for Manhattan, but one street for Paris (both accounting for less than 0.11% and 0.01% of all of the streets), which are distinguished from the redundant others. As a reminder, the "landmark" streets are evaluated from the topological perspective only, and we did not consider any semantic meanings. For example, a very short street can be a landmark if it bears some special meaning. It is important to note that other lines with warm colors such as orange and yellow may constitute part of the mental maps, but they are at the second or third level.

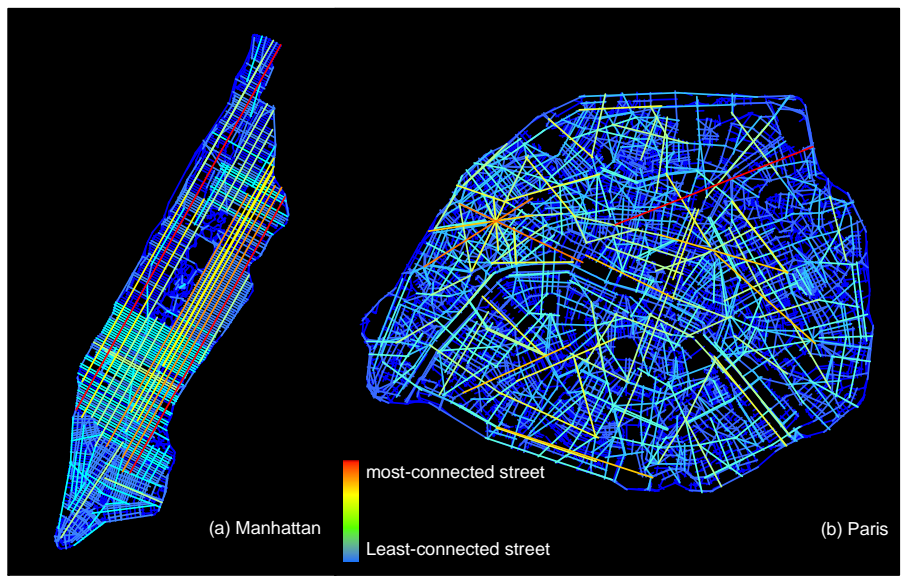

Figure 7: (Color online) Topology of (a) Manhattan and (b) Paris

In this section, we conducted three case studies to illustrate how the image of the city arises from the underlying scaling. We introduced several concepts, such as dead cities versus living cities, surface order versus deep order, and geometry versus topology. These concepts are fundamental to the image of the city. There are some points to note. First, the mental map that we discussed is mainly for the open space of urban environments. Thus, it is from one particular perspective and does not represent a holistic perspective involving all kinds of city artifacts. This kind of simplification is acceptable for an illustration purpose. Second, the mental map is constructed from the city map rather than from the city itself. This is because we rely on the city map for the cognitive mapping processes. Third, neither three-dimensional information nor semantic information was used for the cognitive mapping processes. Note that the increasing availability of geographic information can turn the city map into a highly detailed map containing all related geometric, topological and semantic information of city artifacts or locations. This would help compute the image of the city from a holistic perspective.

**5. Discussions on the information view and the scaling view**
Why can the image of the city emerge in human minds? This is the basic question that we seek to answer in this paper. Haken and Portugali (2003) developed an information view to address this question, i.e. *"the face of the city is its information."* In other words, the city elements that carry more information tend to be more legible and imageable. Eventually, these city elements constitute the image of the city. Or alternatively, cities that carry more information tend to be more legible and imageable than those that carry less information. In this section, we further discuss the scaling view and compare it with the information view.

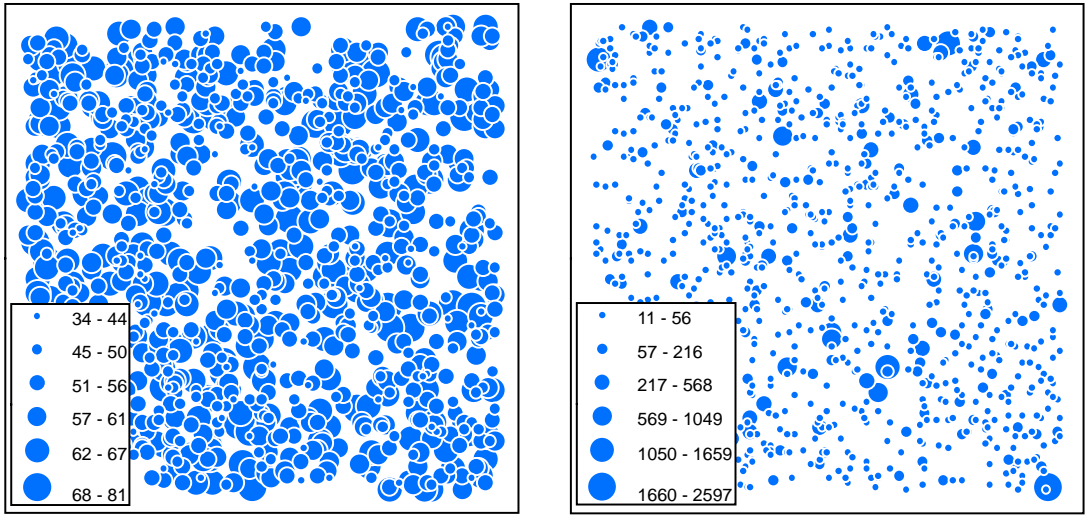

Figure 8: Two point patterns generated from (a) the normal distribution dataset and (b) the heavy-tailed distribution dataset

We adopt the two datasets introduced in Section 3 and generate two point patterns by giving each value (the dot size) a random location as shown Figure 8. Both patterns look arbitrary because the point locations are randomly assigned. However, we claim that the right pattern contains more information than the left one because the amount of information or the information content is about variance (Miller 1956). Apparently, variance of the dot sizes for the right pattern is far larger than that for the left pattern. One can predict a dot size for the left pattern (around the average size 56), while it is far difficult to predict a dot size for the right one. Imagining that one is navigating in the two patterns or landscapes, the right one gives him/her a full surprise from time to time, while the left one is very monotonies due to a lack of changes. This full surprise is in fact the information content. Let us look at how information is measured (Shannon and Weaver 1949). It is measured by

$$H = -\sum_{j=1}^{N} p_j \log_2 p_j,  \qquad [6]$$

where $p_i$ is the probability or frequency of each class and $N$ is the number of classes.

According to Equation (6) and with respect to Table 1, the amount of information for the left pattern is calculated as follows:

$$H_1 = -[(0.069 \times \log_2 0.069) + (0.191 \times \log_2 0.191) + (0.278 \times \log_2 0.278) + (0.245 \times \log_2 0.245) + (0.167 \times \log_2 0.167) + (0.05 \times \log_2 0.05)]$$
$$= 2.38$$

whereas the amount of information for the right pattern is

$$H_2 = -[(0.823 \times \log_2 0.823) + (0.135 \times \log_2 0.135) + (0.028 \times \log_2 0.028) + (0.009 \times \log_2 0.009) + (0.004 \times \log_2 0.004) + (0.001 \times \log_2 0.001)]$$
$$= 0.87$$

The calculation indicates that the left pattern contains far more information than the right pattern, which is incorrect. This result shows a limitation of Shanonian information theory, and justifies Haken's semantic information (Haken 1983). In fact, the left pattern shows little variance because of the normal distribution of dot sizes. Cognitively or perceptually, there is only one class for the left pattern. Thus, there is zero information ($\log_2^1 = 0$). On the other hand, there are at least two perceptible classes: large dots in the head and small dots in the tail. Apparently, the small dots constitute a ground from which a minority of large dots, as a figure, can be distinguished. This perceptible pattern consisting of two classes sets a clear difference from the left pattern that is perceived as one rather homogenous class. Physicists call a normal distribution an even distribution because the variance is too small to be perceived. It is in this sense that the right pattern contains more information than the left pattern. It is also in this sense that the scaling view is consistent with the information view. The scaling view is more intuitive than the information view, since the latter involves calculation of the amount of information. Scaling implies heterogeneity, which indicates more variance and thus more information.

Let us focus on the right pattern to determine how much information the two classes embody. The objects in the head are a minority (e.g., 17.7%), while the objects in the tail are a majority (e.g., 82.3%). According to Equation [6], we can see that the minority of objects in the head contains more information ($0.177 \times \log_2 0.177 = 0.44$), while the majority of objects in the tail contains less information ($0.823 \times \log_2 0.823 = 0.23$). Again, we see how the scaling view is consistent with the information view. The scaling view implies that the city objects in the head (a minority of objects) embody more information than the city objects in the tail. However, the total information for two classes is $0.44 + 0.23 = 0.67$, which is a rough approximation of the information for six classes as shown above (0.87).

The discussion about the two point patterns can be simply applied to cities. If all buildings in a city are similar, there is little variance and thus little information. If on the other hand all buildings are different from each other, they are unlikely to form a coherent pattern as well in human minds due to a lack of grouping. What is unique for the scaling pattern is, a vast majority of buildings (e.g., 90%) are more or less similar, forming a kind of ground; On the top of the ground, a minority of buildings (e.g., 10%) are distinguished to form a kind of figure. The ground-figure contrast is emerged from the head-tail contrast of the scaling. Importantly the contrast or variance due to their enormous imbalance (90/10) can be captured by human minds, forming the image of the city. This is essentially very different from the small variance as shown in Figure 8a, in which the six classes might be recognizable by a computer but not by human minds. It is in this sense that the image of the city arises from the underlying scaling. It is of importance to note that the scaling is the first cause, while the internal representation is the second cause of generating the image of the city. In other words, if a city or its artifacts lacked of the scaling property in the first place, it would not be able to generate a mental map.

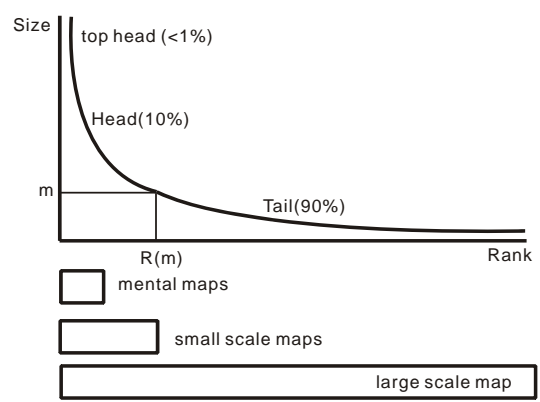

Figure 9: Unification of cognitive and cartographic mapping under the same framework of scaling
(Note: m is the mean of some attribute, while R(m) is the corresponding rank of the mean m)

Not only cognitive mapping, but also cartographic mapping can be governed by the underlying scaling (Jiang 2012a, Jiang et al. 2012). Cognitive mapping is a process composed of a series of psychological transformations by which an individual acquires, codes, stores, recalls, and decodes information about the relative locations and attributes of phenomena in the everyday spatial environment (Downs and Stea 1973). These transformations or mental processes are similar to cartographic processes or geoinformatics, which deal with the acquisition, storage, processing, production, presentation and dissemination of geoinformation in a computer environment. The two seemingly different mapping processes are similar from the point of view of information processing because both are aimed at obtaining an image or a map, either in human minds or in computers, about cities or spatial environment in general. What seems to be a major difference between cognitive and cartographic mapping is that the former is achieved by human minds/brains and the latter by computers. Here, we claim that both cognitive mapping and cartographic mapping can be unified under the same framework of scaling.

The scaling of geographic space is a good thing for mapping because it can guarantee that the small-scale maps have the same patterns or structure as the corresponding large-scale maps. Every time that we select geographic objects in the head or eliminate objects in the tail, the fundamental patterns (or structure) are retained. This selection or elimination is achieved by ranking all geographic objects in a decreasing order and according to their importance or relevance in terms of geometric, topological and even semantic properties (or their combinations). The objects that remain in the top head, or those largest, constitute the content of the cognitive maps. The objects that remain in the head, or those larger ones, constitute the content of the small-scale maps, whereas those in both the head and the tail constitute the content of the large-scale maps. These concepts are illustrated in Figure 9. Note that the head and the tail, or the heavy-tailed distribution, should be understood in a recursive manner; the head can be repeatedly re-distributed or re-stretched as a heavy-tailed distribution. The scaling is such fundamental that it might be formulated as a law of geography or of cartography in particular.

**6. Conclusion**
This paper developed a novel view to explain why the image of the city can be generated from the perspective of cities or city artifacts. This view postulates that the image of the city arises from the underlying scaling of city artifacts or locations, thus leading to a quantitative approach to cognitive mapping. Importantly, we claim that the scaling property is the first cause, while the internal representation is the second cause of generating the image of the city. With the increasing availability of massive amounts of geographic information about cities, in particular large amounts of georeferenced information arising from social media such as Flickr and Twitter, all city artifacts or locations can be ranked, and eventually those largest constitute the image of the city. This ranking process can be based on geometric, topological and/or semantic properties of city artifacts or locations. Thus, we unified cognitive mapping and cartographic mapping under the same framework of scaling. The scaling provides an effective means to compare cities or city artifacts in terms of their legibility and imageability. We relaxed the definition of scaling to include the power law, lognormal and exponential distributions for mapping purposes. We further elaborated that the scaling view is consistent with the information view (Haken and Portugali, 2003), but our scaling view is more intuitive and touches some fundamental aspects of cognitive and cartographic mapping.

This paper is inspired by the work of Alexander (2004) on living structures and how people perceive a structure as pleasant, comfortable and beautiful. Scaling is one of the living structures. The idea of living structures was further developed by the urban theorist Salingaros, whose recent two books (Salingaros 2005, 2006), written in an accessible fashion, have integrated many scientific concepts from mathematics and physics, such as complexity theory and fractal geometry, to theorize architecture and urban design. We believe that the fundamental component of the theories is the underlying scaling present in many effective architectural and city artifacts or urban systems that have living structures or substructures. In this regard, this paper makes a first effort to verify Alexander's architectural theory and philosophy using massive amounts of geographic information about cities. However, many questions remain, such as how to characterize the structural difference between European and North American cities and how to verify the human subconscious linkage to a living structure.

**Acknowledgement**
The author would like to thank XXX for his constructive comments that significantly help improve the quality of this paper.

**References:**
Alexander C. (2004), *The Nature of Order: An essay on the art of building and the nature of the universe*, CES Publishing: CA Berkley.
Barenblatt G. I. (2003), *Scaling*, Cambridge University Press: Cambridge.
Carvalho R and Penn A. (2004), Scaling and universality in the micro-structure of urban space, *Physica A*, 332 539-547.

Clauset A., Shalizi C. R., and Newman M. E. J. (2009), Power-law distributions in empirical data, *SIAM Review*, 51, 661-703.
Downs R. M. and Stea D. (1973), *Image and Environment: Cognitive Mapping and Spatial Behavior*, Aldine Publishing: Chicago.
Gärling T. and Golledge R. G. (1993), *Behavior and Environment: Psychological and geographical approaches*, North-Holland: Amsterdam.
Gibson J. J. (1979), *The Ecological Approach to Visual Perception*, Houghton Mifflin: Boston.
Golledge R. G. and Stimson R. J. (1997), *Spatial Behavior: A geographic perspective*, Guilford Press: New York.
Haken H. (1983), *Synergetics: an introduction*, Springer: Heidelberg.
Haken H. and Portugali J. (2003), The face of the city is its information, *Journal of Environmental Psychology*, 23, 385 – 408.
Hanson J. (1989), Order and structure in urban design: the plans for the rebuilding of London after The Great Fire of 1666, *Ekistics*, 56(334-335), 22-42.
Hillier B. (1996), *Space Is the Machine: a configurational theory of architecture*, Cambridge University Press: Cambridge.
Hillier B. and Hanson J. (1984), *The Social Logic of Space*, Cambridge University Press: Cambridge.
Holland J. H. (1995), *Hidden Order: How adaption builds complexity*, Perseus Books: New York.
Jacobs A. B. (1995), *Great Streets*, MIT Press: Cambridge, MA.
Jenks G. F. (1967), The data model concept in statistical mapping, *International Yearbook of Cartography*, 7, 186-190.
Jiang B. (2007), A topological pattern of urban street networks: universality and peculiarity, *Physica A: Statistical Mechanics and its Applications*, 384, 647 - 655.
Jiang B. (2009), Street hierarchies: a minority of streets account for a majority of traffic flow, *International Journal of Geographical Information Science*, 23(8), 1033-1048.
Jiang B. (2012a), Head/tail breaks: A new classification scheme for data with a heavy-tailed distribution, *The Professional Geographer*, xx(x), xx – xx.
Jiang B. (2012b), Computing the image of the city, In: Campagna M., De Montis A., Isola F., Lai S., Pira C. and Zoppi C. (editors, 2012), *Planning Support Tools: Policy analysis, implementation and evaluation, Proceedings of the 7th Int. conf. on Informatics and Urban and Regional Planning INPUT 2012*, 111-121.
Jiang B. and Liu X. (2012), Scaling of geographic space from the perspective of city and field blocks and using volunteered geographic information, *International Journal of Geographical Information Science*, 26(2), 215-229.
Jiang B., Liu X. and Jia T. (2012), Scaling of geographic space as a universal rule for map generalization, Preprint: http://arxiv.org/abs/1102.1561
Kitchin R. and Freundschuh S. (editors, 2000), *Cognitive Mapping: Past, Present and Future*, Routledge: London.
Lynch K. (1960), *The Image of the City*, The MIT Press: Cambridge, Massachusetts.
Miller G. A. (1956), The magic number seven, plus or minus two: Some limits on our capacity for processing information, *The Psychological Review*, 63(2), 81–97.
Montello D. R. (2002), Cognitive map-design research in the twentieth century: Theoretical and empirical approaches, *Cartography and Geographic Information Science*, 29, 283-304.
Montello D. R. (2009), Cognitive research in GIScience: Recent achievements and future prospects. *Geography Compass*, 3(5), 1824–1840.
Omer I., Goldblatt R. and Or U. (2005), Virtual city design based on urban image theory, *The Cartographic Journal*, 42(1), 1-12.
Penn A. (2003), Space syntax and spatial cognition: Or, why the axial line? *Environment and Behavior*, 35, 30-65.
Portugali J. (editor, 1996), *The Construction of Cognitive Maps*, Springer: Berlin.
Salingaros N. A. (2005), *Principles of Urban Structure*, Techne: Delft.
Salingaros N. A. (2006), *A Theory of Architecture*, UMBAU-VERLAG: Solingen.
Shannon C. E. and Weaver W. (1949), *The Mathematical Theory of Communication*, University of Illinois Press: Urbana, IL.
Surowiecki J. (2005), *The Wisdom of Crowds: Why the many are smarter than the few and how collective wisdom shapes business, economies, societies and nations*, Abacus: London.
Yarbus A. L. (1967), *Eye Movements and Vision*, Plenum Press: New York.
Zipf G. K. (1949), *Human Behaviour and the Principles of Least Effort*, Addison Wesley: Cambridge, MA.